\documentclass[aps,pra,twocolumn,groupedaddress]{revtex4}

\usepackage[latin1]{inputenc}
\usepackage[T1]{fontenc}
\usepackage[english]{babel}
\usepackage[dvips]{graphicx}
\usepackage{enumerate,amsthm,amsmath,amssymb,color,graphicx,bbm,float}
\usepackage{natbib}

\newtheorem{theo}{Theorem} 

\usepackage{epstopdf}

\begin{document}

\newcommand{\ket} [1] {\vert #1 \rangle}
\newcommand{\bra} [1] {\langle #1 \vert}
\newcommand{\braket}[2]{\langle #1 | #2 \rangle}
\newcommand{\proj}[1]{\ket{#1}\bra{#1}}
\newcommand{\mean}[1]{\langle #1 \rangle}
\newcommand{\opnorm}[1]{|\!|\!|#1|\!|\!|_2}

\title{Percolation of secret correlations in a network}

\author{Anthony Leverrier}
\affiliation{ICFO-Institut de Ci\`encies Fot\`oniques, 08860 Castelldefels (Barcelona), Spain}

\author{Ra\'{u}l Garc{\'i}a-Patr\'{o}n}
\affiliation{Research Laboratory of Electronics, MIT, Cambridge, MA 02139}
\affiliation{Max-Planck Institut fur Quantenoptik, Hans-Kopfermann Str. 1, D-85748 Garching, Germany}

\date{\today} 

\begin{abstract}
In this work, we explore the analogy between entanglement and secret 
classical correlations in the context of large networks, more precisely
the question of percolation of secret correlations in a network. 
It is known that entanglement percolation in quantum networks can display a
highly nontrivial behavior depending on the topology of the
network and on the presence of entanglement between the nodes.
Here we show that this behavior, thought to be of a genuine
quantum nature, also occurs in a classical context. 
\end{abstract}

\maketitle

In 1993, Maurer introduced an information-theoretically secure secret-key agreement
scenario where two honest parties, Alice and Bob, have access
to many independent outcomes of random variables $A,B$ correlated
with the eavesdropper's (Eve) variable $E$ through the probability distribution
$P_{A,B,E}(a,b,e)$. 
Their goal is to extract a secret key from their data
with the help of; (i) local manipulations of their respective variables,
using protocols such as error correction codes and privacy amplification;
(ii) communicating over a public channel, \emph{i.e.},
using local operations and public communication \cite{mau93}.

It was later observed in \cite{CP02,CMS02} 
that Maurer's scenario shares a lot of similitudes with the
quantum scenario where Alice, Bob and Eve share an initial quantum state
$\rho_{ABE}$ and Alice and Bob's task is to distill a maximum amount
of entanglement qubits (ebits), \emph{i.e.},
\begin{equation}
 \ket{\psi}_{AB}=\frac{1}{\sqrt{2}}[\ket{00}+\ket{11}],
\end{equation}
using local operations and classical communication. 
In the same way as entanglement can be seen as a resource that cannot increase under 
local operations and classical communication, 
secret classical correlations are measured in secret bits (sbits), \emph{i.e.},
\begin{equation}
\label{sbit}
P_{A,B,E}(a,b,e) = \frac{1}{2}\delta_{a,b} P_E(e),
\end{equation}
a universal resource that cannot increase under local operations
and public communication.
In this expression, $\delta_{a,b} = 1$ if $a=b$ and $\delta_{a,b} =0$ otherwise and $P_E(e)$ refers to any possible distribution of Eve's random variable $e$, which is therefore completely uncorrelated with Alice and Bob's variables. 
In \cite{CP02}, it was shown that many quantum information
processing protocols have an equivalent protocol in Maurer's
secure secret-key scenario. For example,
the analog of quantum teleportation 
is simply one-time pad, see Figure 1 (b). 
Similarly, entanglement distillation, entanglement dilution,
(probabilistic) single-copy conversion were also shown to have
secure secret-key analogous protocols. It is not surprising then
that entanglement measures, such as the \textit{entanglement distillation}
and \textit{entanglement of formation}, have their corresponding
secure secret-key measure \cite{mau93,GW00,RW03}.
The connection between entanglement and secure 
secret-key has benefited the research in both fields. On the first hand,
Gisin and Wolf asked whether a classical secrecy analog 
of bound entanglement \cite{GW00} existed. This question was 
positively answered in \cite{ACM04}, where a tripartite (plus Eve)
distribution $P_{ABCE}(a,b,c,e)$ was shown to need previously established
secrecy between the honest parties to be generated,
but from which no secret key could be distilled. 
Despite further results \cite{MA06, PB11}, it still an open
question whether there exists bipartite bound-secrecy 
while bipartite bound-entanglement is known to exist.
On the other hand, the secrecy measure \textit{intrinsic information}, introduced in
\cite{MW99} and shown to be a lower-bound of the \textit{secret distillation}
and an upper-bound of the \textit{secret of formation} in \cite{RW03},
was generalized to the quantum scenario in \cite{CW04}. There, the authors
introduced the \textit{squashed entanglement} 
measure which has recently received a lot of attention \cite{BCY10,CSW10}.

In this paper, we want to explore the analogy between entanglement and secret 
classical correlations in the context of large networks. More precisely,
we study the percolation of secret correlations in lattices. 
In the quantum case, when the goal is to establish ebits between two 
arbitrary nodes of a quantum lattice, there exists a phase transition 
for entanglement percolation for which the success probability does not 
decrease exponentially with the distance between the two nodes \cite{ACL07}.
More interestingly, Ref. \cite{ACL07} gave the first example of a quantum protocol that changes the topology of the network,
 making possible the distillation of a perfect entanglement link in a regime
where traditional percolation would fail. 
This phenomenon was further studied in \cite{PCA08,LWL09, CC10, PCL10, DB10} and 
extended to the mixed state scenario \cite{BDJ09,per10,BDJ10}. 
In the present work, we show that the same phenomenon already happens
in the purely classical context of Maurer's  
secret-key agreement scenario.

\section{Secret-Key Networks}
\label{model}

\begin{figure}[!h!]
\begin{center}
\includegraphics[width=1\linewidth]{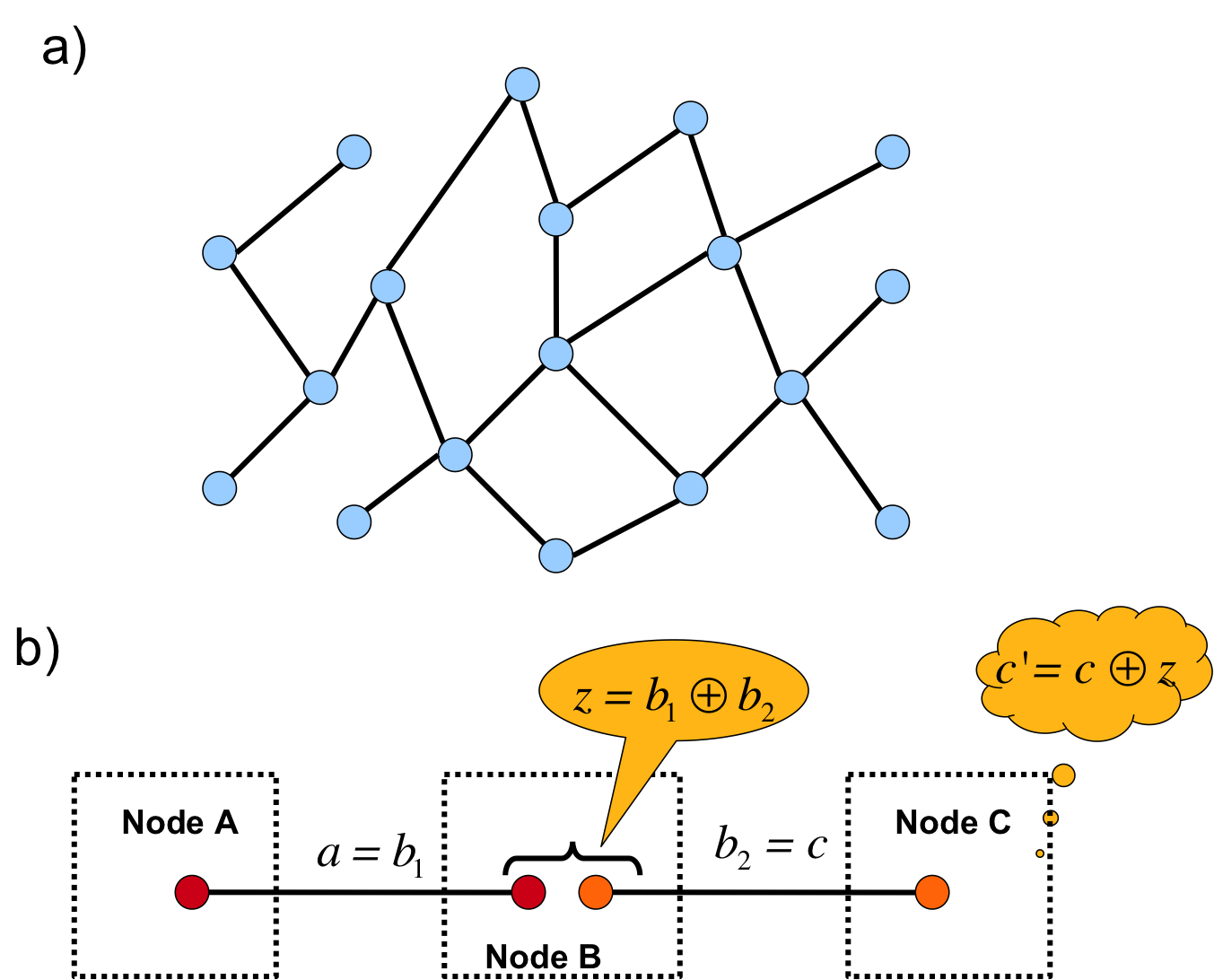}
\end{center}
\caption{(Color online.) Secret-key networks: a) A general secret-key network is
composed of a set of nodes (vertices of the graph) distributed with a given geometry,
sharing secrecy correlations when connected by a link (edges of the graph).
b) \textit{One-time Pad}: In order to establish a unbiased secret bit between, 
previously unconnected, nodes $A$ and $C$, $B$ and $C$ apply one round of one-time pad;
(i) $B$ publicly announces the value of $z=b_1\oplus b_2$; (ii)
$C$ calculates $c'=c\oplus z$ which gives $c'=a$.}
\label{Fig:network}
\end{figure}

In this work, we study secrecy distribution in secret-key networks, see Fig.\ref{Fig:network}, (a). 
More precisely, we are interested in secret-key networks where each edge 
$\bar{ab}$, between nodes $A$ and $B$, corresponds to a biased secret-key bit
\begin{equation}
\label{biased}
P_{A,B,E}(a,b,e) = [(1-p)\delta_{a,b,0} + p\delta_{a,b,1}]*P_E(e)
\end{equation}
where $\delta_{a,b,x} := \delta_{ab} \delta_{ax}$ and $p \leq 1/2$.

The question we want to address is the following: given a secret-key network and 
a choice of two nodes,
does there exist a strategy, based on local manipulations of the bits and public classical
communication, allowing to distill a secret bit (sbit) between these two nodes? 
Let us first start by considering some simple examples of networks as their analysis 
will be useful for the rest of the paper.

\subsection{Simple examples}

\emph{A single link}: The simplest network 
consists of two nodes, $A$ and $B$, sharing a biased secret bit 
following a Bernoulli distribution of parameter $p \leq 1/2$
given by Eq. (\ref{biased}). The results for
probabilistic conversion of \cite{CP02} (see Appendix \ref{basics}) show that 
the probability to convert this biased secret bit into an unbiased
one is equal to $2p$.
A protocol achieving this optimal value is the following. Let $a$ be Alice and Bob's bit. 
If $a=0$ (which happens with probability $1-p \geq 1/2$), Alice tosses a biased coin that 
gives "heads" with probability $(1-2p)/(1-p)$. If she gets "heads", she tells Bob to 
abort the protocol; otherwise, they keep $a$ as the final sbit. It is easy to 
check that conditioned on the fact that the protocol did not abort, the value of 
$a$ is unbiased. 

\emph{A chain with 2 links}: consider the scenario shown in Fig. \ref{Fig:network}, (b)
with three nodes where $A$ and $B$ share a biased 
secret bit ($a=b_1$) while $B$ and $C$ share a second biased secret bit $b_2=c$.
The probability of establishing a unbiased bit
between nodes $A$ and $C$ can only be lower or equal to the probability
of conversion of a single link. Surprisingly, there exists a strategy
succeeding with average probability $2p$. This strategy uses
one-time pad, the secret-key protocol analogous to quantum teleportation: 
node $B$ simply publicly announces the value of $b_1\oplus b_2$. If
$b_1\oplus b_2 = 1$, which happens with probability $2p(1-p)$, $C$ flips his bit and obtains an unbiased secret bit shared with $A$.
If $b_1 \oplus b_2=0$, $A$ and $C$ secret-key (unnormalized) distribution becomes
\begin{equation}
P_{A,C,E}(a,c,e)\propto [(1-p)^2\delta_{a,c,0}+p^2\delta_{a,c,1}]*P_E(e),
\end{equation}
which has a conversion probability $P_c=2\frac{p^2}{p^2+(1-p)^2}$.
Putting everything together gives an average probability of success of 
\begin{eqnarray}
 P_{succ}&=&P(b_1\oplus b_2 = 1)*1+P(b_1\oplus b_2 = 0)*P_c \nonumber \\
&=&2p(1-p)+2p^2=2p.
\end{eqnarray}

\emph{Two parallel links}: if the nodes $A$ and $B$ share two biased secret bits $a_1$ and $a_2$ (with $p \leq 1-1/\sqrt{2}$), the optimal probabilistic conversion strategy (see Theorem 2 of Appendix 1) consists for nodes $A$ and $B$ in mapping their two bits 
into a new bit $a_f$  such that $a_f=0$ if $a_1=a_2=0$ and $a_f=1$ otherwise. 
The bit $a_f$ then follows a Bernoulli distribution 
with parameter $(1-p)^2$ and the probability to convert it into an unbiased 
secret bit is $2(1-(1-p)^2) = 2p(2-p)$.

\subsection{The straightforward strategy}
\label{straightforward}

As in the quantum scenario \cite{ACL07}, there exists one natural strategy to distill 
an unbiased secret bit between two arbitrary nodes, $A$ and $B$, 
of a given lattice $\mathbbm{L}$. This protocol consists in trying to convert each 
biased secret bit (corresponding to each edge of the lattice) into an unbiased secret bit, each 
conversion succeeding with some probability $p_\mathrm{succ}$. 
If there exists a path among the edges of the unbiased secret bit graph
connecting nodes $A$ and $B$, then, using 
one-time pad along this path, one can produce a secret bit
between nodes $A$ and $B$. Based on percolation theory, one can show that
the probability that two arbitrary nodes are connected by a path does not depend on their distance
in the graph if $p_\mathrm{succ}$ is larger than the
critical percolation threshold probability $p_c^{\mathbbm{L}}$ of the lattice.
For  $p_\mathrm{succ}\leq p_c^{\mathbbm{L}}$
the success probability of the overall procedure decreases exponentially 
with the distance in the lattice between the two nodes (see Appendix \ref{percolation} for details).
The question that one wishes to answer is whether or not this simple strategy is optimal 
and whether the bound corresponding to $p_c^{\mathbbm{L}}$ is tight. In the case of entanglement percolation,
it was shown in \cite{ACL07} 
that the strategy described above is asymptotically optimal in the case of one-dimensional 
chains but not in general for two-dimensional lattices. 
In the following, we show that these two statements also apply to the case of 
secret classical correlations. 

\section{One-dimensional chain}
\label{1d}

\subsection{Presentation of the problem}

\begin{figure}[!h!]
\begin{center}
\includegraphics[width=1\linewidth]{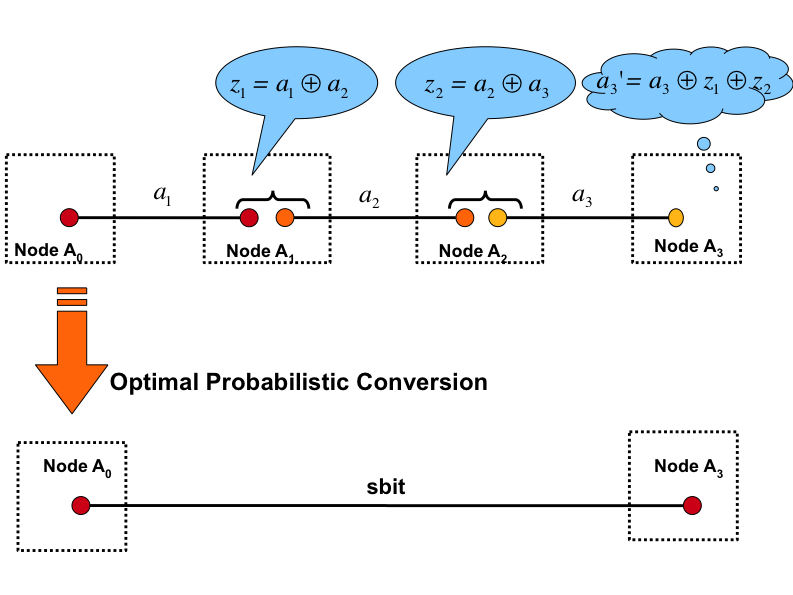}
\end{center}
\caption{(Color online.) 1D-chain (4 nodes in the figure): 
The protocol giving the best probability of distilling a secret bit between
nodes $A_0$ and $A_n$ works as follows: 
(i) Each intermediate node, $A_i$ ($1\leq i\leq n-1$), 
makes public the value of the sum (modulo 2) of its two bits: $z_i=a_{i}\oplus a_{i+1}$;
(ii) Node $A_n$ calculates privatly $a_n\oplus \sum_{i=1}^{n-1}z_i$;
(iii) Nodes $A_0$ and $A_n$ apply the optimal probabilistic conversion protocol
for the given set ($z_1,...,z_{n-1}$).}
\label{Fig:1D_chain}
\end{figure}

Let us consider a one-dimensional chain with $n$ links and
$n+1$ nodes: $A_0, A_1, \cdots, A_n$. Each link $i$ corresponds to a pair of
biased perfectly correlated variables, as in Eq.(\ref{biased}).
Because each pair is perfectly correlated, we simplify the discussion
by noting $a_i$ the single bit shared by $A_{i-1}$ and $A_i$, as shown
in Fig. \ref{Fig:1D_chain}.
In this model, the eavesdropper has no prior information on the bits $a_i$ (except for the value of $p$), meaning that her
initial probability distribution is uncorrelated with $(a_1, \cdots, a_n)$.
We will now show that the probability of establishing a 
perfect secret bit between the extremities of a chain, 
(between $A_0$ and $A_n$) decreases exponentially fast with $n$, 
except if the chain is initially composed of perfect secret bits, \emph{i.e}.,
if $p= 1/2$. Hence, with that respect, distribution of secrecy and distribution 
of entanglement display the same behavior in the case of one-dimensional chains.

\subsection{Description of the optimal protocol}
\label{only_protocol}
 
Let us first start with the smallest, but non-trivial, case of 2 links. 
The general proof will then follow by induction.
As shown on Fig. \ref{Fig:1D_chain}, nodes $A_0$ and $A_1$ share the biased secret bit $a_1$ and nodes $A_1$ 
and $A_2$ share $a_2$. 
Both bits, $a_1$ and $a_2$, are biased and have value 0 with probability $1-p$. 
The goal is for $A_0$ and $A_2$ is to distill a secret bit
unknown to Eve, who had no prior information on $a_1$ and $a_2$. 

In order to succeed, node $A_1$ has to publicly announce some information, $z_1$, depending on his own bits $a_1$ and $a_2$ and possibly on some random ancillary
bits. This public information should allow nodes $A_0$ and $A_2$
to distill a secret bit, without giving any information to Eve. 
In full generality, node $A_1$ may use a probabilistic strategy to generate $z_1$. 
However, because every probabilistic
strategy is a convex combination of deterministic ones, a probabilistic strategy cannot be better than the best deterministic
one. Therefore, it is sufficient to consider the
set of deterministic functions $z_1=f(a_1,a_2)$. 
Since $A_1$ simply needs to tell node $A_2$
whether it should keep its bit $a_{2}$ or  flip it 
in order to match the secret bit $a_1$, 
$z_1$ only needs to take 2 possible values, 0 or 1. As a consequence, we only
need to analyse 16 possible functions $f$ of $a_1$ and $a_2$. 
The constraints of the problem help us find the only possibility for $f$.
First, node $A_2$ should be able to recover $a_1$ from the knowledge of $a_2$ and $z_1$, 
imposing $f(0, a_2) \ne f(1, a_2)$. Second, 
Eve should not learn any information about $a_1$, 
imposing $\sum_{a_2} f(0, a_2) = \sum_{a_2} f(1,a_2)$.
Up to a relabeling, the only function that satisfies these constraints is the 
\emph{exclusive or} (XOR): $f(a_1, a_2) = a_1 \oplus a_2$. It is not surprising
that we obtain exactly the one-time pad protocol, which achieves
a success probability of $2p$ for a three node chain, as shown before.

\emph{Generalization to $n$ links}:  
In a scenario with more links, it is easy to see that the same 
reasoning applies. In particular, all the intermediate nodes should announce 
the XOR of their two bits, up to some relabeling.  
The protocol is therefore the following. Each intermediate node $A_i$ 
(for $1 \leq i \leq n-1$) publicly announces $z_i = a_{i}\oplus a_{i+1}$. 
The final node can then compute the value of $a_1$ since 
$a_1 = a_n \oplus z_{n-1} \oplus z_{n-2} \oplus \cdots \oplus z_1$. 
Once nodes $A_0$ and $A_n$ share this (biased) secret bit, 
they can proceed with the optimal probabilistic conversion protocol of 
Theorem \ref{prob_conversion} and end up with a unbiased secret bit. 
The average success probability reads
\begin{equation}
p_n = \sum_{z_1, \cdots, z_{n-1}} p(z_1, \cdots, z_{n-1}) p(\mathrm{success}|z_1, \cdots, z_{n-1})
\end{equation}
where $p(\mathrm{success}|z_1, \cdots, z_{n-1})$ corresponds to the success 
probability of conversion of the bit given that the vector announced by 
the intermediate nodes is $(z_1, \cdots, z_{n-1})$.
The success probability is equal to (twice) the first half of the binomial expansion (see Appendix \ref{analysis}): 
\begin{equation}
p_n =  2 \sum_{\text{first half}} {n \choose k} p^{n-k}(1-p)^k 
\end{equation}
which can be lower bounded, as shown in the Appendix \ref{analysis}, by
\begin{equation}
p_n \leq (2 \sqrt{p(1-p)})^{n}.
\end{equation}
Despite doing better than the straighforward strategy,
which gives a success probability of $(2p)^n$,
it still decreases exponentially fast with $n$ for initial
unbiased secret bit ($p \ne 1/2$), similarly as in
the quantum scenario \cite{ACL07}.
This result is not surprising as for percolation to occur, 
it is crucial that the topology of the 
network allows for many different paths between two given nodes to exist,
which is not possible in a one-dimensional chain.

\section{Two-dimensional case}
\label{2d}

A possible strategy to distill a secret-key
between arbitrary nodes of a lattice consists in using the straightforward
strategy presented in Section \ref{straightforward}. First, one tries to 
distill a secret bit over each link 
in the lattice. If the probability of success is above the 
percolation threshold of the lattice, then with a positive probability, 
this procedure created a path consisting of secret bits 
between the two arbitrary nodes. This path can then be used to
establish a secret bit between these two nodes thanks to
one-time pad. 

Following Ref.\cite{ACL07}, we now give an explicit example of a 
2-dimensional lattice where local preprocessing and public communication
allows one to change the topology of the initial lattice into another one with a lower percolation threshold.  
This shows that there exist non-trivial strategies that
succeed to establish a secret bit between two nodes
when the naive strategy would fail.
The protocol involves three basic operations 
that were described in Section \ref{model}:
(i) the conversion of a single link into an sbit with success probability $p_1=2p$; 
(ii) the conversion of two consecutive links into an sbit, 
also with success probability $p_2=2p$;
(iii) the conversion of two parallel links, with success probability $p_{//} = 2p(2-p)$.

\begin{figure}[!h!]
\begin{center}
\includegraphics[width=1\linewidth]{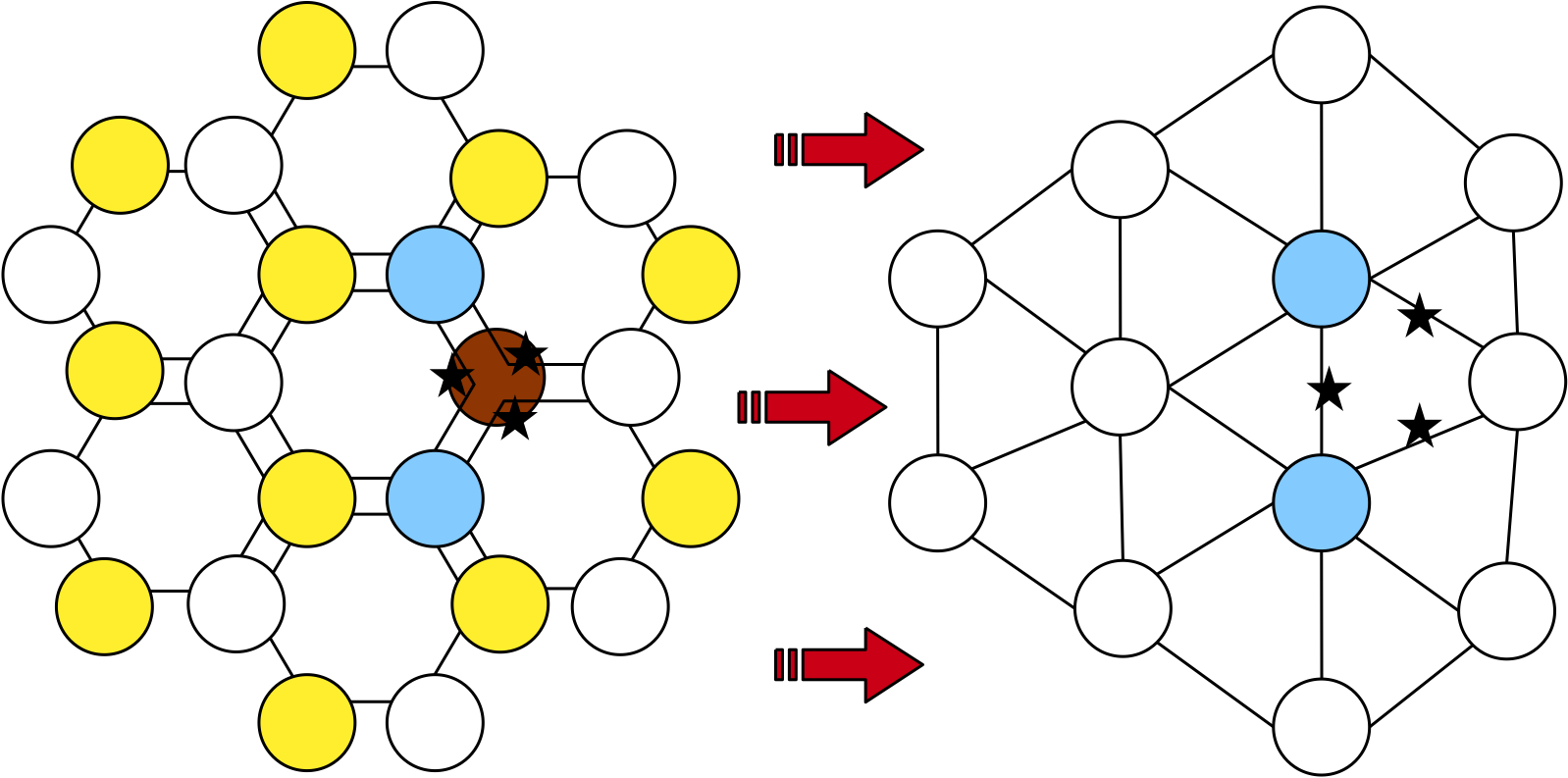}
\end{center}
\caption{(Color online.) The initial configuration is a honeycomb (hexagonal) lattice,
each node being represented by a circle. Each two adjaccent
nodes are connected by two links, each one representing a biased
secret-key bit. Using local operations and public communication,
we can tranform the topology to a triangular lattice improving the percolation
threshold. To do so, every \textit{yellow} node performs three parallel one-time pad operations,
one for each pair of connected links (labeled by a star in the interior of the \textit{brown} node).
This tranforms every pair of connected links into a link of the
new triangular lattice (see black stars for those corresponding to the brown node).} 
\label{lattice}
\end{figure}

Let us consider an initial honeycomb (hexagonal) lattice, as shown in Figure \ref{lattice}, 
where each link of the lattice consists of 2 biased secret-key bits with parameter $p$. 
For this lattice, the naive strategy of Section \ref{straightforward} succeeds with constant probability 
as soon as $p_{//} \geq p_c^{\mathrm{hex}}$, where the 
percolation threshold probability for the honeycomb lattice is given by
\begin{equation}
p_c^{\mathrm{hex}} = 1 - \sin(2\pi/18) \approx 0.6527.
\end{equation}
Therefore, the straightforward percolation
strategy succeeds only for $p \geq 0.1792$.

As illustrated in Ref. \cite{ACL07}, one might consider a more elaborate strategy, 
namely, one can try to change the topology of the lattice in order to facilitate percolation.
As shown in Figure 3, the idea is that half the nodes from the original lattice should work 
together in order to create a triangular lattice. To do that, 
these nodes perform one-time pad over each pair of connected biased secret-key bits.
It is easy to see that each link of the triangular lattice will be distilled into 
an sbit with probability $p_2 = 2p$. Percolation can then occur as soon as $p_2$ 
exceeds the threshold percolation probability of the triangular lattice,
\begin{equation}
 p_c^{\mathrm{triang}}  =  \sin(2\pi/18) \approx 0.3473.
\end{equation}
This "topology conversion" strategy is therefore compatible with percolation 
of  sbits for $p \geq 0.1736$.
We conclude that in the regime where $p\in [0.1736, 0.1792 ]$, percolation 
can occur if the nodes use the non-trivial percolation strategy 
consisting in changing the topology of the lattice from honeycomb to triangular, 
while the straightforward strategy fails. 
Other quantum percolation examples \cite{PCA08,LWL09}
can also been easily adapted to the secet-key percolation
scenario, using the tools presented here.

\section{Conclusion}

In this paper, using known analogies between entanglement and classical
secret-key correlations, we have studied secrecy percolation in networks. 
More precisely, we have shown that local operations
and public communication can be used to change the topology of
a secrecy network and to establish a secret key between nodes, in a regime where the initial lattice configuration is not compatible with percolation of secrecy. 
This effect was already known to exist in quantum entanglement networks. 
Our work shows that this phenomenon thought to be of a genuine 
quantum nature, already appears in the context of classical secret correlations.

\section*{Acknowledgments}

The authors acknowledge fruitful discussions with 
Antonio Ac\'in. A. L. received financial support from the EU ERC Starting grant PERCENT. 
R. G.-P. acknowledges financial support from the W. M. Keck Foundation
Center for Extreme Quantum Information Theory and the Alexander von Humboldt foundation.

\appendix

\section*{Appendix}
\subsection{Pure State Conversions}
\label{basics}

In Ref. \cite{CP02}, the authors characterized the set of transformations 
which are allowed among probability distributions. Their characterization 
is reminiscent of the quantum case \cite{Nie99} and uses the same notion of majorization.
\begin{theo}[Deterministic conversion \cite{CP02}]
\label{determ_conversion}
If Alice and Bob begin with an arbitrary classical bipartite pure state,
$P_{ABE}(i, j, k) = \delta_{i,j} p_i P_E(k)$, then they can produce 
a new state $P_{ABE}'(i, j, k) = \delta_{i,j} q_i P_E(k)$ if and only if $\vec{q}$ majorizes $\vec{p}$.
\end{theo}

Recall that the vector $\vec{q} = \left\{ q_i\right\}$ is said to majorize 
the vector $\vec{p} = \left\{ p_i\right\}$ (with $p_1 \geq p_2 \geq \cdots$ 
and $q_1 \geq q_2 \geq \cdots$) if
\begin{equation}
\sum_{i=1}^k q_i \geq \sum_{i=1}^k p_i \quad \forall k.
\end{equation} 

Whereas Theorem \ref{determ_conversion} is only concerned 
with conversion strategies which work with probability 1, 
the following result deals with strategies which work with 
a finite probability. Note that again, one recovers the 
same result as in the quantum case \cite{Vid99}.
\begin{theo}[Probabilistic conversion \cite{CP02}]
\label{prob_conversion}
If Alice and Bob begin with an arbitrary classical bipartite pure state, $P_{ABE}(i, j, k) = \delta_{i,j} p_i P_E(k)$, then the maximal probability with which they can produce a new state $P_{ABE}'(i, j, k) = \delta_{i,j} q_i P_E(k)$ is given by
\begin{equation}
\min_{k} \frac{1-\sum_{i=1}^k p_i}{1-\sum_{i=1}^k q_i}.
\end{equation}
\end{theo}

\subsection{Bond-percolation in Lattices}
\label{percolation}

The percolation behaviors that appear in the context of quantum networks or secrecy networks are closely related to the concept of bond-percolation. 
The scenario of bond-percolation is the following. Consider a lattice $\mathbb{L}$ such that for each edge of $\mathbb{L}$, the bond is open (or equivalently, the edge is present) with probability $p$. Taking the limit where the size of $\mathbb{L}$ is infinite, one can define the probability $\theta(p)$ that a randomly chosen node belongs to a cluster of infinite size. Then, there exists a critical percolation probability $p_c^{\mathbb{L}}$ such that:
\begin{itemize}
\item $\theta(p) > 0$ if $p > p_c^\mathbbm{L}$,
\item $\theta(p) = 0$ if $p < p_c^\mathbbm{L}$. 
\end{itemize}
The link to our problem is immediate. Given two arbitrary nodes of the lattice, one is interested in whether an unbiased secret bit can be established between them. In the case where there exists an infinite size component, then both nodes belong to this cluster with probability $\theta^2(p)$ and an unbiased secret bit can be established between them. Otherwise, if there is no cluster of infinite size, the probability of establishing an unbiased secret bit decreases exponentially with the distance between the nodes in the lattice $\mathbbm{L}$. 

\subsection{Analysis of the protocol of Section \ref{only_protocol}}
\label{analysis}
 
Let us consider the same scenario of a chain of $n$ links where each link is  
a biased secret bit that takes value 1 with probability $p \leq 1/2$, 
and bound the probability of creating a secret bit between the extremities.
 
As we saw in Section \ref{only_protocol}, the protocol consists first in 
publicly announcing the vector ${\bf z}= (z_1, \cdots, z_{n-1})$, and then conditionally on the 
value of ${\bf z}$, try to convert the bit shared by $A_0$ and $A_n$ into an sbit. 
The probability of success $p_n$ of this procedure is therefore given by:
\begin{equation}
p_n = \sum_{z_1, \cdots, z_{n-1}} p(z_1, \cdots, z_{n-1}) p(\mathrm{success}|z_1, \cdots, z_{n-1})
\end{equation}
where $p(\mathrm{success}|z_1, \cdots, z_{n-1})$ corresponds to the success 
probability of conversion of the bit given that the vector announced by 
the intermediate nodes is $(z_1, \cdots, z_{n-1})$.

The probability that the public communication is 
described by ${\bf z} = (z_1, \cdots, z_{n-1})$ is
\begin{eqnarray*}
p({\bf z} )& =& p(a_1 = 0, a_2 = z_1,  \cdots, a_{n}= \bigoplus_k z_k) + \\
&&p(a_1 = 1, a_2 = 1 \oplus z_1, \cdots, a_{n}= 1  \bigoplus_k z_k)\\
\end{eqnarray*}

Given a particular value of ${\bf z}$, the success probability 
for the probabilistic conversion of Theorem \ref{prob_conversion} reads:
\begin{multline}
p(\mathrm{success}|{\bf z}) = \\
\frac{\min (p(a_1 = 0,  \cdots, a_{n}= \bigoplus z_k) ,
p(a_1 = 1,  \cdots, a_{n}= 1  \bigoplus z_k)) }{p(a_1 = 0,  \cdots, a_{n}= \bigoplus_k z_k) +
p(a_1 = 1,  \cdots, a_{n}= 1  \bigoplus z_k)}.\nonumber
 \end{multline} 

Putting everything together, one has
\begin{eqnarray*}
p_n &=& 2 \sum_{{\bf a}} \min (p(a_1,\cdots, a_{n}), p(\overline{a_1}, \cdots, \overline{a_{n}}))\\
&=& 2 \sum_{{\bf a}} \min (p^{w({\bf a})}(1-p)^{n-w({\bf a})}, p^{n-w({\bf a})}(1-p)^{w({\bf a})})\\
&=&2 \sum_{{\bf a}}  p^{n-w({\bf a})}(1-p)^{w({\bf a})}\\
&=& 2 \sum_{\text{first half}} {n \choose k} p^{n-k}(1-p)^k
\end{eqnarray*}
where ${\bf a} =(a_1, \cdots, a_{n})$, $\overline{a_k} := 1 \oplus a_k$, $w({\bf a})$ 
denotes the Hamming weight of the vector ${\bf a}$ and "first half" 
means that the sum contains exactly the first half of the binomial expansion, 
that is, the $2^{n-1}$ first terms of this expansion.

This probability is achieved if all the intermediate nodes 
($n-1$ such nodes) reveal the value of the $XOR$ of their two bits.

This success probability is equal to (twice) the first half of the binomial expansion. 
Let us bound this quantity:
\begin{eqnarray}
p_n &=&  2 \sum_{\text{first half}}{n \choose k} p^{n-k}(1-p)^k \\
& \leq & 2 p^{\lfloor n/2 \rfloor}(1-p)^{\lceil n/2 \rceil}  \sum_{\text{first half}}{n \choose k} \\
& \leq &  p^{\lfloor n/2 \rfloor}(1-p)^{\lceil n/2 \rceil}  \sum_{k=0}^n {n \choose k} \\
& \leq & 2^n p^{\lfloor n/2 \rfloor}(1-p)^{\lceil n/2 \rceil} \\
& \leq & (2 \sqrt{p(1-p)})^{n},
\end{eqnarray}
which goes down to 0 exponentially fast with n for $p \ne 1/2$.


\begin{thebibliography}{23}
\expandafter\ifx\csname natexlab\endcsname\relax\def\natexlab#1{#1}\fi
\expandafter\ifx\csname bibnamefont\endcsname\relax
  \def\bibnamefont#1{#1}\fi
\expandafter\ifx\csname bibfnamefont\endcsname\relax
  \def\bibfnamefont#1{#1}\fi
\expandafter\ifx\csname citenamefont\endcsname\relax
  \def\citenamefont#1{#1}\fi
\expandafter\ifx\csname url\endcsname\relax
  \def\url#1{\texttt{#1}}\fi
\expandafter\ifx\csname urlprefix\endcsname\relax\def\urlprefix{URL }\fi
\providecommand{\bibinfo}[2]{#2}
\providecommand{\eprint}[2][]{\url{#2}}

\bibitem[{\citenamefont{Maurer}(1993)}]{mau93}
\bibinfo{author}{\bibfnamefont{U.~M.} \bibnamefont{Maurer}},
  \bibinfo{journal}{Information Theory, IEEE Transactions on}
  \textbf{\bibinfo{volume}{39}}, \bibinfo{pages}{733} (\bibinfo{year}{1993}).

\bibitem[{\citenamefont{Collins and Popescu}(2002)}]{CP02}
\bibinfo{author}{\bibfnamefont{D.}~\bibnamefont{Collins}} \bibnamefont{and}
  \bibinfo{author}{\bibfnamefont{S.}~\bibnamefont{Popescu}},
  \bibinfo{journal}{Phys. Rev. A} \textbf{\bibinfo{volume}{65}},
  \bibinfo{pages}{32321} (\bibinfo{year}{2002}).

\bibitem[{\citenamefont{Cerf et~al.}(2002)\citenamefont{Cerf, Massar, and
  Schneider}}]{CMS02}
\bibinfo{author}{\bibfnamefont{N.~J.} \bibnamefont{Cerf}},
  \bibinfo{author}{\bibfnamefont{S.}~\bibnamefont{Massar}}, \bibnamefont{and}
  \bibinfo{author}{\bibfnamefont{S.}~\bibnamefont{Schneider}},
  \bibinfo{journal}{Phys. Rev. A} \textbf{\bibinfo{volume}{66}},
  \bibinfo{pages}{042309} (\bibinfo{year}{2002}).

\bibitem[{\citenamefont{Gisin and Wolf}(2000)}]{GW00}
\bibinfo{author}{\bibfnamefont{N.}~\bibnamefont{Gisin}} \bibnamefont{and}
  \bibinfo{author}{\bibfnamefont{S.}~\bibnamefont{Wolf}}, in
  \emph{\bibinfo{booktitle}{Advances in Cryptology, CRYPTO 2000}}
  (\bibinfo{organization}{Springer}, \bibinfo{year}{2000}), pp.
  \bibinfo{pages}{482--500}.

\bibitem[{\citenamefont{Renner and Wolf}(2003)}]{RW03}
\bibinfo{author}{\bibfnamefont{R.}~\bibnamefont{Renner}} \bibnamefont{and}
  \bibinfo{author}{\bibfnamefont{S.}~\bibnamefont{Wolf}}, pp.
  \bibinfo{pages}{562--577} (\bibinfo{year}{2003}).

\bibitem[{\citenamefont{Ac{\'i}n et~al.}(2004)\citenamefont{Ac{\'i}n, Cirac,
  and Masanes}}]{ACM04}
\bibinfo{author}{\bibfnamefont{A.}~\bibnamefont{Ac{\'i}n}},
  \bibinfo{author}{\bibfnamefont{J.~I.} \bibnamefont{Cirac}}, \bibnamefont{and}
  \bibinfo{author}{\bibfnamefont{L.}~\bibnamefont{Masanes}},
  \bibinfo{journal}{Phys. Rev. Lett.} \textbf{\bibinfo{volume}{92}},
  \bibinfo{pages}{107903} (\bibinfo{year}{2004}).

\bibitem[{\citenamefont{Masanes and Ac{\'i}n}(2006)}]{MA06}
\bibinfo{author}{\bibfnamefont{L.}~\bibnamefont{Masanes}} \bibnamefont{and}
  \bibinfo{author}{\bibfnamefont{A.}~\bibnamefont{Ac{\'i}n}},
  \bibinfo{journal}{Information Theory, IEEE Transactions on}
  \textbf{\bibinfo{volume}{52}}, \bibinfo{pages}{4686} (\bibinfo{year}{2006}).

\bibitem[{\citenamefont{Prettico and Bae}(2011)}]{PB11}
\bibinfo{author}{\bibfnamefont{G.}~\bibnamefont{Prettico}} \bibnamefont{and}
  \bibinfo{author}{\bibfnamefont{J.}~\bibnamefont{Bae}},
  \bibinfo{journal}{Phys. Rev. A} \textbf{\bibinfo{volume}{83}},
  \bibinfo{pages}{042336} (\bibinfo{year}{2011}).

\bibitem[{\citenamefont{Maurer and Wolf}(1999)}]{MW99}
\bibinfo{author}{\bibfnamefont{U.~M.} \bibnamefont{Maurer}} \bibnamefont{and}
  \bibinfo{author}{\bibfnamefont{S.}~\bibnamefont{Wolf}},
  \bibinfo{journal}{Information Theory, IEEE Transactions on}
  \textbf{\bibinfo{volume}{45}}, \bibinfo{pages}{499} (\bibinfo{year}{1999}).

\bibitem[{\citenamefont{Christandl and Winter}(2004)}]{CW04}
\bibinfo{author}{\bibfnamefont{M.}~\bibnamefont{Christandl}} \bibnamefont{and}
  \bibinfo{author}{\bibfnamefont{A.}~\bibnamefont{Winter}},
  \bibinfo{journal}{Journal of mathematical physics}
  \textbf{\bibinfo{volume}{45}}, \bibinfo{pages}{829} (\bibinfo{year}{2004}).

\bibitem[{\citenamefont{Brand{\~{a}}o et~al.}(2010)\citenamefont{Brand{\~{a}}o,
  Christandl, and Yard}}]{BCY10}
\bibinfo{author}{\bibfnamefont{F.~G. S.~L.} \bibnamefont{Brand{\~{a}}o}},
  \bibinfo{author}{\bibfnamefont{M.}~\bibnamefont{Christandl}},
  \bibnamefont{and} \bibinfo{author}{\bibfnamefont{J.}~\bibnamefont{Yard}},
  \bibinfo{journal}{Arxiv preprint arXiv:1010.1750}  (\bibinfo{year}{2010}).

\bibitem[{\citenamefont{Christandl et~al.}(2010)\citenamefont{Christandl,
  Schuch, and Winter}}]{CSW10}
\bibinfo{author}{\bibfnamefont{M.}~\bibnamefont{Christandl}},
  \bibinfo{author}{\bibfnamefont{N.}~\bibnamefont{Schuch}}, \bibnamefont{and}
  \bibinfo{author}{\bibfnamefont{A.}~\bibnamefont{Winter}},
  \bibinfo{journal}{Phys. Rev. Lett.} \textbf{\bibinfo{volume}{104}},
  \bibinfo{pages}{240405} (\bibinfo{year}{2010}).

\bibitem[{\citenamefont{Ac{\'i}n et~al.}(2007)\citenamefont{Ac{\'i}n, Cirac,
  and Lewenstein}}]{ACL07}
\bibinfo{author}{\bibfnamefont{A.}~\bibnamefont{Ac{\'i}n}},
  \bibinfo{author}{\bibfnamefont{J.}~\bibnamefont{Cirac}}, \bibnamefont{and}
  \bibinfo{author}{\bibfnamefont{M.}~\bibnamefont{Lewenstein}},
  \bibinfo{journal}{Nature Phys.} \textbf{\bibinfo{volume}{3}},
  \bibinfo{pages}{256} (\bibinfo{year}{2007}).

\bibitem[{\citenamefont{Perseguers et~al.}(2008)\citenamefont{Perseguers,
  Cirac, Ac\'in, Lewenstein, and Wehr}}]{PCA08}
\bibinfo{author}{\bibfnamefont{S.}~\bibnamefont{Perseguers}},
  \bibinfo{author}{\bibfnamefont{J.~I.} \bibnamefont{Cirac}},
  \bibinfo{author}{\bibfnamefont{A.}~\bibnamefont{Ac\'in}},
  \bibinfo{author}{\bibfnamefont{M.}~\bibnamefont{Lewenstein}},
  \bibnamefont{and} \bibinfo{author}{\bibfnamefont{J.}~\bibnamefont{Wehr}},
  \bibinfo{journal}{Phys. Rev. A} \textbf{\bibinfo{volume}{77}},
  \bibinfo{pages}{022308} (\bibinfo{year}{2008}).

\bibitem[{\citenamefont{Lapeyre et~al.}(2009)\citenamefont{Lapeyre, Wehr, and
  Lewenstein}}]{LWL09}
\bibinfo{author}{\bibfnamefont{G.~J.} \bibnamefont{Lapeyre}},
  \bibinfo{author}{\bibfnamefont{J.}~\bibnamefont{Wehr}}, \bibnamefont{and}
  \bibinfo{author}{\bibfnamefont{M.}~\bibnamefont{Lewenstein}},
  \bibinfo{journal}{Phys. Rev. A} \textbf{\bibinfo{volume}{79}},
  \bibinfo{pages}{042324} (\bibinfo{year}{2009}).

\bibitem[{\citenamefont{Cuquet and Calsamiglia}(2009)}]{CC10}
\bibinfo{author}{\bibfnamefont{M.}~\bibnamefont{Cuquet}} \bibnamefont{and}
  \bibinfo{author}{\bibfnamefont{J.}~\bibnamefont{Calsamiglia}},
  \bibinfo{journal}{Phys. Rev. Lett.} \textbf{\bibinfo{volume}{103}},
  \bibinfo{pages}{240503} (\bibinfo{year}{2009}).

\bibitem[{\citenamefont{Perseguers et~al.}(2010)\citenamefont{Perseguers,
  Cavalcanti, Lapeyre, Lewenstein, and Ac{\'i}n}}]{PCL10}
\bibinfo{author}{\bibfnamefont{S.}~\bibnamefont{Perseguers}},
  \bibinfo{author}{\bibfnamefont{D.}~\bibnamefont{Cavalcanti}},
  \bibinfo{author}{\bibfnamefont{G.~J.} \bibnamefont{Lapeyre}},
  \bibinfo{author}{\bibfnamefont{M.}~\bibnamefont{Lewenstein}},
  \bibnamefont{and} \bibinfo{author}{\bibfnamefont{A.}~\bibnamefont{Ac{\'i}n}},
  \bibinfo{journal}{Phys. Rev. A} \textbf{\bibinfo{volume}{81}},
  \bibinfo{pages}{032327} (\bibinfo{year}{2010}).

\bibitem[{\citenamefont{Di~Franco and Ballester}(2010)}]{DB10}
\bibinfo{author}{\bibfnamefont{C.}~\bibnamefont{Di~Franco}} \bibnamefont{and}
  \bibinfo{author}{\bibfnamefont{D.}~\bibnamefont{Ballester}},
  \bibinfo{journal}{Arxiv preprint arXiv:1008.1679}  (\bibinfo{year}{2010}).

\bibitem[{\citenamefont{Broadfoot et~al.}(2009)\citenamefont{Broadfoot, Dorner,
  and Jaksch}}]{BDJ09}
\bibinfo{author}{\bibfnamefont{S.}~\bibnamefont{Broadfoot}},
  \bibinfo{author}{\bibfnamefont{U.}~\bibnamefont{Dorner}}, \bibnamefont{and}
  \bibinfo{author}{\bibfnamefont{D.}~\bibnamefont{Jaksch}},
  \bibinfo{journal}{EPL (Europhysics Letters)} \textbf{\bibinfo{volume}{88}},
  \bibinfo{pages}{50002} (\bibinfo{year}{2009}).

\bibitem[{\citenamefont{Perseguers}(2010)}]{per10}
\bibinfo{author}{\bibfnamefont{S.}~\bibnamefont{Perseguers}},
  \bibinfo{journal}{Phys. Rev. A} \textbf{\bibinfo{volume}{81}},
  \bibinfo{pages}{012310} (\bibinfo{year}{2010}).

\bibitem[{\citenamefont{Broadfoot et~al.}(2010)\citenamefont{Broadfoot, Dorner,
  and Jaksch}}]{BDJ10}
\bibinfo{author}{\bibfnamefont{S.}~\bibnamefont{Broadfoot}},
  \bibinfo{author}{\bibfnamefont{U.}~\bibnamefont{Dorner}}, \bibnamefont{and}
  \bibinfo{author}{\bibfnamefont{D.}~\bibnamefont{Jaksch}},
  \bibinfo{journal}{Phys. Rev. A} \textbf{\bibinfo{volume}{82}},
  \bibinfo{pages}{042326} (\bibinfo{year}{2010}).

\bibitem[{\citenamefont{Nielsen}(1999)}]{Nie99}
\bibinfo{author}{\bibfnamefont{M.~A.} \bibnamefont{Nielsen}},
  \bibinfo{journal}{Phys. Rev. Lett.} \textbf{\bibinfo{volume}{83}},
  \bibinfo{pages}{436} (\bibinfo{year}{1999}).

\bibitem[{\citenamefont{Vidal}(1999)}]{Vid99}
\bibinfo{author}{\bibfnamefont{G.}~\bibnamefont{Vidal}},
  \bibinfo{journal}{Phys. Rev. Lett.} \textbf{\bibinfo{volume}{83}},
  \bibinfo{pages}{1046} (\bibinfo{year}{1999}).

\end{thebibliography}

\end{document}